\documentstyle[12pt,aaspp]{article}
\tighten

\received{1994 April 8}
\revised{DATE}
\accepted{1994 June 20}
\journalid{VOL}{JOURNAL DATE}
\articleid{Start}{Finish}
\paperid{Manuscript id}
\lefthead{BAND}
\righthead{GRB Time Dilation}

\begin{document}
\title{COSMOLOGICAL TIME DILATION \\ IN GAMMA RAY BURSTS?}
\author{David L. Band}
\affil{CASS, University of California, San Diego, CA  92093 \\
Internet: dlbbat@cass09.ucsd.edu \\
{\it Received 1994 April 8; accepted 1994 June 20} \\
To appear in {\it The Astrophysical Journal Letters} }

\begin{abstract}
Norris et al. (1994) report that the temporal structure of faint gamma ray
bursts is longer than that of bright bursts, as expected for time dilation
in the cosmological models of burst origin.  I show that the observed trends
can easily be produced by a burst luminosity function and thus may not result
from cosmological effects.  A cosmological signature may be present, but the
tests Norris et al. present are not powerful enough to detect these signatures.
\end{abstract}
\section{Introduction}
Cosmological models for the origin of gamma ray bursts are based solely on the
intensity and spatial distributions of the BATSE bursts (e.g., Meegan et al.
1992), and not on any intrinsic burst features.  Therefore, Norris and his
collaborators (Norris et al. 1993a,b, 1994; Norris 1994; Davis et al. 1994)
are searching for evidence of cosmological time dilation in
burst time histories.  If bursts have a cosmological origin then
the most distant (and therefore faintest) bursts BATSE detects
originate at $z\sim 1$ (e.g., Wickramasinghe et al. 1993) and their time
structure should be stretched by a factor of $\sim2$ compared to nearby
bursts.
Norris and collaborators present a series of tests which probe burst
temporal structures on different time scales, and report a
consistent dilation of $\sim 2$ between the faint and intense bursts.
Given the importance of this result, it
is crucial that the tests be evaluated; here I focus on the tests presented
by Norris et al. (1994; hereafter N94).

N94 assume the peak photon emission rate (i.e., the maximum photon luminosity)
is constant, and therefore the peak flux is an exact distance measure
(R.~Nemiroff 1994, personal communication).  This assumption is overly
simplistic.  Bursts display a broad range of spectral shapes (Band et al.
1993), and the fraction of the total emission observed within a given spectral
band varies.  One can argue on physical grounds that the total energy release
(e.g., a neutron star binding energy) or peak energy luminosity (e.g., an
Eddington luminosity, or a sharply peaked luminosity function such as proposed
by Mao \& Yi
1994) should be constant; however, with the variety of burst
spectra and durations observed the peak photon luminosity in a particular
energy band will most likely not be constant.  The peak count rate has become
the paramount measure of burst intensity not because of burst physics, but
because detectors trigger on it (e.g., Paczy\'nski \& Long 1988).  I will show
that when the ``standard candle'' assumption is relaxed, the burst correlations
which N94 state are consistent with a cosmological origin can easily be
products of the burst luminosity function.

N94's search for a cosmological signature entails extensive transformations of
the burst time histories.  To give all the bursts in their sample the same
signal-to-noise ratio and apparent distance, N94 normalize the bursts so that
they have the same peak flux, by necessity that of the weakest burst in their
sample, and add the appropriate Poisson noise.  N94 justify the degradation of
the high quality data from bright bursts as necessary to remove biases
resulting from differing signal-to-noise ratios.
In the various tests they present, N94 use only BATSE bursts with durations
greater than 1.5s.  They consider averages for three burst groups
characterized by peak flux:  bright, dim and dimmest.  Each sample has
approximately 40 bursts.  For each burst 65.536s of data was used; the basic
data had a time resolution of 64ms, although for some purposes the resolution
was reduced.

N94 present three tests which are intended to probe different time scales.  I
analyze each test in turn (\S 2-4), and then end with some conclusions (\S 5).
\section{Test 1---Normalized Count Fluences}
N94 calculate the average number of counts above background in the
normalized time histories.  This normalized count fluence is effectively
the observed photon fluence $S_N$ divided by peak photon flux $P$,
a quantity with units of time which can be interpreted as a duration.
Assuming a cosmological burst origin, we expect $S_N/P\propto 1+z$,
and indeed the average normalized photon fluence is about twice as large
for the two dim burst samples as for the bright burst group.  While
this test is physically sensible, operationally it compares a function
of the peak flux ($S_N/P$) to the peak flux itself, a procedure which
can give a spurious result.  For example, if $S_N$ and $P$
are uncorrelated, then clearly $S_N/P$ will be inversely correlated with $P$.
While $S_N$ and $P$ should be correlated because they both decrease with source
distance, there are more subtle non-cosmological effects which can produce
this apparent signature.

The cosmological effects result from time dilation and spectral redshifting
which are continuous functions of the redshift, affecting even the bright
bursts.  If bursts have
a power law spectrum $N(E)\propto E^\alpha$ over the energy band of interest,
then the redshift dependence is $P\propto (1+z)^\alpha /d_m^2$ and
$S_N\propto (1+z)^{1+\alpha} /d_m^2$, where $d_m$ is the proper motion
distance.  Thus $S_N/P \propto 1+z$, as expected.  Assuming $q=1/2$ (i.e.,
an $\Omega=1$ cosmology) and $\alpha=-1$ then
$P= [(\sqrt{1+z_m}-1)/(\sqrt{1+z}-1)]^2$, where $P$ is scaled by its
value at $z_m\sim 1$.

Figure~1 shows $\langle S_N/P \rangle$ as a function of $P$ for both
cosmological bursts with constant peak photon luminosity $L$ (related to $P$)
and for sources distributed uniformly in Euclidean space out to a maximum
radius.  For the Euclidean case I use $n(L)\propto L$ for $L\le L_0$ and
$n(L)\propto L^{-3}$ for $L\ge L_0$ as the luminosity function, chosen so
that the number of bursts converges.  Note that 85\% of the bursts fall within
a luminosity range of 6.5, consistent with the constraints of Horack et al.
(1994).  As long as the total number of photons $N$ emitted (related
to $S_N$) is uncorrelated with $L$, its distribution is irrelevant to relative
values of $\langle S_N/P \rangle$.  The maximum $z_m=1.25$ used by N94 and a
burst spectral index $\alpha=-1$ are assumed.  For the Euclidean model I
normalize $\langle S_N/P \rangle$ to 1 at $P=100$.  In
the cosmological model the bright bursts
show some time dilation and the difference in dilation between the dim and
dimmest bursts is significant.  In the Euclidean model the bursts at any given
value of $P$ originate from a range of distances (unless $L$ is a standard
candle, in which case a single distance is singled out), with the more distant
bursts providing, on average, smaller values of $S_N$.  However, if the source
density decreases beyond some distance, there are fewer low $S_N$ bursts as
$P$ decreases, raising the average $S_N$.  Thus a Euclidean model can have
correlations similar to cosmological time dilation.

N94 assume that the photon luminosity $L$ is a standard candle, while in the
example in Figure~1 I assume that $L$ is not constant and that the
photon emission $N$ is uncorrelated with $L$.  These two cases are drawn from
a continuum of possible distributions of $L$ and $N$.  Both the joint
distribution of $L$ and $N$ (as also shown by Brainerd 1994)
and the geometry of the source distribution affect
the observed correlation of $\langle S_N /P \rangle$ with $P$.  In a Euclidean
source geometry the dependence of $\langle S_N/P \rangle$ on $P$ is determined
by the relation between $L$ and $\langle N \rangle_L$ ($N$ averaged at constant
$L$): if flatter than $\langle N \rangle_L \propto L$ then
$\langle S_N /P\rangle$ increases at small $P$.  The extent of the change of
$\langle S_N/P \rangle$ at small $P$ is a function of the dynamic range of $L$
and the abruptness of the source distribution's spatial cutoff.  Consequently
the distributions of $N$ and $P$ in Euclidean space can very easily mimic a
cosmological signature, and thus this test is inconclusive.
The second and third tests in Davis et al.
(1994) are essentially the same as the first test in N94.
\vfill\eject
\section{Test 2---Wavelets}
Wavelets find a signal's frequency content as a
function of time.  N94 use the Haar transform which, while
not optimal in temporally resolving the frequency content (Daubechies 1992,
pp. 10-13), can nonetheless be calculated easily.  N94
calculate Haar transforms on time scales ranging from
512ms to 65.536s, a total of 8 time scales.  They then average the absolute
value of the transform amplitudes on each time scale, producing an ``activity''
at discrete time scales
which is akin to the power spectrum for the traditional Fourier transform.
The activities for all the bursts of a brightness sample are then
averaged.  For time scales less than $\sim 2$s the
average activity is flat at the same value for each burst sample, as
expected for Poisson noise (Poisson noise is ``white'').  On longer
time scales the average activity curve increases with time scale.  However,
the curves get progressively flatter as the sample gets brighter; thus the
dimmest sample lies above, and the bright sample below, the dim sample.  N94
compare simulations of dilated and undilated burst distributions and find that
the undilated average activity curve lies below the dilated curve; once again
the observations appear to be consistent with cosmological bursts.  However,
N94 do not develop a detailed quantitative or qualitative understanding of
how cosmological time dilation produces the activity curves, e.g., in terms of
the widths of the spikes within the bursts, the separation between spikes, the
number of counts in a burst,
etc.  In addition, the simulations do not calibrate the magnitude of the
cosmological effect.

Investigating the Haar transform analytically provides greater insight into
the activity curves.  The Haar wavelet transform
of the function $f(x)$, using a time scale hierarchy separated by
factors of 2, is
\begin{equation}
T_{m,n} = \int dx\,f(x) 2^{-m/2} \Psi\left(2^{-m}x-n\right),
   \quad m\ge 0, \quad  n\ge 0
\end{equation}
where $\Psi(y)=1$ for $0\le y \le 1/2$, -1 for $1/2< y \le 1$, and 0 otherwise.
The time variable $x$ is in units of the shortest time scale, and thus data
points are every half unit.  Finally, $m$ indicates the time scale and $n$ the
time interval.  N94 define the
activity as the average transform for a given time scale, or
\begin{equation}
A_m = \langle T_m \rangle = {1\over J} \sum_{n=0}^{J-1}  |T_{m,n}|
\end{equation}
where there are $J$ time intervals characterized by the time scale indexed by
$m$.  The activity is defined at only a small number of discrete
time scales; line segments are drawn in to guide the eye.  If $f(x)=a$ for
half the basic time scale (i.e., a single data point), and zero otherwise
(i.e., effectively a $\delta$-function for
the time scales under consideration), the activity for a data window whose
length is $2^N$ of the shortest time scales (with $2^{N+1}$ data points) is
\begin{equation}
A_m = 2^{1/2-N+m/2} a \quad , \quad m\ge 0.
\end{equation}
The dependence on $N$ results from averaging the single nonzero element over
the entire data window.  Thus on successively longer time scales (each twice
the length of the preceding time scale) the activity increases by a factor of
$2^{1/2}$ (as seen from the $m$ dependence in eqn.~[3]): the activity curve
(joining the small number of points where the activity is actually calculated)
will be a power law with index $1/2$.  Eqn. (3) can be generalized to signals
greater than half the basic time scale; signals with a duration less than
half a given time scale will appear to be constant on half that time scale in
calculating the activity for
longer time scales.  The activity of a $\delta$-function is not flat because
the activity was defined as the average of $|T_{m,n}|$ and not of $T_{m,n}^2$.
On the other hand, white noise has a flat activity curve.  Each amplitude
$T_{m,n}$ for the transform of noise has approximately the same magnitude, so
the averages of its absolute value and its square are both constant.
Therefore the burst activity curve has three regions:  a flat, noise-dominated
region at short time scales; intermediate times scales in which burst
structures are evident; and time scales longer than twice the burst duration
where the activity curve is a power law with index $1/2$ and a normalization
proportional to the total number of counts in the burst.

Figure 2 shows the activity of GB910717 observed by BATSE (trigger \#543---see
Fishman et al. 1994 for the time history).  The signal-to-noise ratio is
sufficiently large such that noise may dominate only the shortest time scales
(i.e., only for $m=0$ and 1 does the activity curve appear to be flat).
The burst duration is less than 7s, and thus on time scales of 16.384s
(more than twice the duration) and greater the activity is a power law with
index 1/2 (the linear segment for $m\ge7$).

The activity for a burst which has undergone cosmological time dilation keeping
the peak flux constant (N94 normalize all time histories to the same peak
flux) follows from the relation $F(x)=f(x/(1+z))$ between $f(x)$ and $F(x)$,
the time histories in the rest and observer's frames, respectively:
\begin{equation}
A_{m,obs} = (1+z)^{3/2} A_{[m-\ln_2(1+z)], rest}
\end{equation}
where $\ln_2$ is the base 2 logarithm (i.e., $y=\ln_2 2^y$).  Since structures
are shifted to longer time scales, activity on one time scale in the rest
frame is translated to activity on a longer time scale in the observer's
frame.  On this longer time scale the activity, which averages over the burst
duration and post-burst background, is diluted by less background, an effect
which contributes a factor of $1+z$.  Finally, the transforms have a different
normalization on the new time scale, which provides the last factor of
$(1+z)^{1/2}$.  Thus we expect the time dilated activity curve to be shifted
to longer time scales and up relative to the undilated curve.  Figure~2 also
shows the activity for a time dilated burst at $z=1$ where the photon rate has
been kept constant.  A complication arises when the dilated burst is longer
than the data window since then part of the dilated time history is not
included in the calculation of the activity, and the dilution of burst activity
by post-burst background is reduced.  Because of the reduced dilution, the
leading factor in eqn. (4) ranges from $(1+z)^{1/2}$ up to the maximum of
$(1+z)^{3/2}$.  In addition, the character of the temporal
structure may vary across a burst.
Barring such systematic trends, we expect the dilated activity curve
to have the same shape as the undilated curve, albeit shifted to longer time
scales; the relative normalization depends on the data window.

Thus time dilation should shift the average activity curve to longer
time scales.  The Haar transform used here is calculated for time scales
separated by factors of 2, and therefore the dilation for bursts at
$z=1$ shifts the activity by one time scale.  The bursts in the
simulations of N94 are at $z=1.25$, and therefore the dilated and
undilated activities cannot be related cleanly.  The
simulations in Norris (1994) use a redshift of $z=1$, and we find that
the predicted relationship between activity curves holds approximately
at the large time scales: the dilated activity is shifted
to the next longest time and up by a factor of $\sim 2.1$.  Since the dilated
and undilated samples use different simulated bursts the exact translation
predicted by eqn.~(4) is not expected.  However, the simulated dilated and
undilated curves clearly diverge, whereas the observed activity curves in N94
are not so cleanly separated.  In particular, on the 32.768s
time scale the activity curves for the dim and bright samples intersect.
Therefore the consistency of the observed activity curves with the simulated
curves in N94 is not overwhelmingly convincing.

The first test shows that the average count fluence in the normalized time
histories is twice as large for the dim and dimmest burst samples as for the
bright sample.  Therefore we expect that on the longest time scales, which
will usually be longer than the burst duration, the activities of the dim and
dimmer samples will be larger than for the bright sample:  in eqn.~(3), which
is applicable to time scales longer than twice the burst duration, the
activity is proportional to number of counts.  Thus the separation between dim
and bright bursts may be a natural consequence of the difference in the
normalized photon fluence and not of dilation.
Thus this test may be no more than the first test in a different form.

\section{Test 3---Mitrofanov Peak Alignment}

Pioneered by Mitrofanov et al. (1993), this test aligns the highest peaks of
each normalized burst, and considers the peak width of the average of the
aligned time histories.  The average peak should be dilated for the more
distant bursts if they are at cosmological distances, and indeed N94 find
that the peaks get
successively broader as the bright, dim and dimmest samples are compared.
Simulations of cosmological time dilation produce broadening similar to that
observed in the data, indicating the observations are consistent with
$z\sim 1$ for BATSE's most distant bursts.

Note that Mitrofanov et al. (1994) perform the same test on bursts selected
from the first BATSE catalog (Fishman et al. 1994) with durations ($T_{90}$)
greater than 1s, and do not find a difference between the dim and bright
events.  However, Mitrofanov et al. divide the bursts into broader intensity
groups and include shorter bursts than N94, making time dilation, if present,
more difficult to detect.  The authors of both papers have additional
unpublished results (personal communication from both J.~Norris and
I.~Mitrofanov, 1994), and we must await future resolution of this observational
issue.

I am concerned that in implementing this test N94 may have inadvertantly
broadened some burst structures by filtering their time histories to make
the peaks more evident.  The time histories' wavelet
transforms for each time scale are reduced by the amplitude expected for
noise, and then the time histories are reconstituted by taking the inverse
transform.  Consequently both noise and signal are filtered out on the
shortest time scales.  The wavelet activities used in the second test indicate
that Poisson noise dominates for time scales shorter than $\sim 2$s, of order
of the peak widths in this peak alignment test.  The same filtering was used
for the normalized time histories (with added Poisson noise) of all three
flux groups, and thus all three should be affected similarly.  However, N94
do not provide the extent of the filtering on each time scale (i.e., the
reduction in the wavelet transform for each sample).  Note that this filtering
conserves the number of counts in the time history since the wavelet basis
functions have zero mean.

The peaks of the dim bursts may be broader than the bright bursts because
the faint bursts have more normalized counts, the phenomenon found by the
first test.  Since the peaks of the normalized time histories have the same
value, the dim bursts have more counts by having more emission surrounding
the peak; the averaged time history will most likely have a broader peak.
I showed in \S 2 that low peak flux bursts can have more normalized counts
if the intrinsic peak photon luminosity function is uncorrelated with the total
photon emission; a distribution of peak FWHMs at constant photon emission
will produce such a luminosity function.  Once again, this test's results may
be a consequence of the effect found by the first test.
\section{Conclusions}
The results of the second and third tests in N94 are consequences of the first
test:  if the normalized photon fluence of the dim bursts is greater than that
of the bright bursts, the wavelet ``activity'' on long time scales (which is
proportional to the number of counts) will also be greater, and the averaged
normalized time profiles should be broader (to get more counts under the
time history curve with the same peak flux).  The effect observed in the
first test---dim bursts have more counts when normalized by the peak
flux---can very easily be a spurious correlation if there is a distribution
of peak photon luminosities.  Consequently N94 must demonstrate that
the peak photon luminosity is a standard candle for their tests to be
convincing.

It should be noted that N94 do not compare the observations to quantitative
predictions of the cosmological model.  Only local bursts and bursts at the
maximum redshift, $z_m=1.25$, are simulated.  Yet the redshift
varies across each of the peak flux bins, reducing the observed cosmological
signature.  Also, it is not clear how sensitive the observations are to the
parameters of the cosmological model, e.g., to the value of $z_m$.

Given the importance of determining the burst distance scale, we need tests of
the cosmological hypothesis.  Unfortunately, many of these tests are easily
confounded by very plausible relations among burst properties or by broad
intrinsic distributions.  Thus comparing the observed burst durations and the
peak flux $P$ is more robust than using $S_N/P$ and $P$.  However, the duration
and the peak photon luminosity $L$ could be inversely related, if for example
the total photon emission $N$ is constant.  Also, the cosmological signature
is diluted by a broad peak photon luminosity function.  In addition, an
accurate
determination of the duration is more difficult for fainter bursts; detailed
study of the measured duration as a function of burst intensity is required
for this test (as well as for any systematic use of the duration as a burst
property).  Similarly, searches for the redshifting of burst spectra
(Paczy\'nski 1992; Mao \& Paczy\'nski 1992; Dermer 1992; Brainerd 1993) are
complicated by the diversity of burst spectra (Band et al. 1993) and the
unknown photon luminosity function.  Nonetheless, additional tests of the
cosmological hypothesis should be sought; sophisticated techniques such as
those discussed by Loredo \& Wasserman (1994) are designed to compare
different model distributions to the observations.

As N94 point out, observing the expected cosmological time dilation does not
prove that the burst sources are at cosmological distances since burst physics
may conspire to produce the same properties.  Conversely, showing that the
apparent temporal broadening in faint bursts can easily be a product of burst
property distributions does not prove that the effect is not time dilation.
Observing a
theory's predicted consequence helps validate that theory.  However, the
argument in a theory's favor is less convincing if the predicted effect can
have other plausible causes.  In this study I showed that the tests presented
by N94 could be misleading, and therefore these tests should not increase our
confidence in the validity of the cosmological hypothesis.
\vfill\eject
\acknowledgments{I thank the members of the BATSE team and the high energy
astrophysics group at UCSD with whom I discussed this work.  I also thank
R.~Nemiroff and J.~Norris, the referee, for informative and collegial
discussions concerning their paper (N94).
This research was supported by NASA contract NAS8-36081.}

%
%
\centerline{\bf Figures}

Figure 1.  Model relationships between normalized counts ($S_N/P$) and peak
photon flux $P$.  The peak flux is normalized by $P_t$, the faintest flux
BATSE can detect.  The solid curve is for a cosmological model where BATSE
detects bursts to $z_m=1.25$, and the peak photon luminosity is constant.
The dashed curve is $\langle S_N/P \rangle$ for a homogeneous source
population (in Euclidean space) out to a finite radius $R_M$; the bursts in
this scenario have a broken power law luminosity function $\phi(L)\propto L$
for $L<L_0$ and $\phi(L)\propto L^{-3}$ for $L\ge L_0$ where
$L_0 = 0.75\times 4\pi R_M^2 P_t$.  For the Euclidean case
$\langle S_N/P \rangle$ is normalized to 1 at $P=100$, while for the
cosmological case $\langle S_N/P \rangle$ is normalized by its asymptotic
large $P$ value.  The vertical lines indicate the sample bins used by N94:
dimmest ($P=1$ to 1.71); dim ($P=1.71$ to 3.21); and bright ($P=12.9$ to 178).

Figure 2.  Wavelet activity for burst GB910717 (solid curve), assumed to be
local, and for the same burst dilated to $z=1$ maintaining the same count rate
(dashed curve).  The activity
is the wavelet analog to a Fourier power spectrum; it is defined at a small
number of discrete time scales.  Note that the dashed curve is the same as the
solid curve shifted to one larger time scale and up by $2^{3/2}$.

\end{document}